\documentclass[12pt]{article}
\usepackage[cp1250]{inputenc}
\usepackage{amsmath}
\usepackage{amssymb}

\newtheorem{Theorem}{Theorem}[section]

\newtheorem{Lemma}[Theorem]{Lemma}
\newtheorem{Definition}[Theorem]{Definition}
\newenvironment{Proof}[1]{{\bf Proof #1.} }{$\Box$\\}

\newcommand{\vn}{\vec{n}}
\newcommand{\vtau}{\vec{\tau}}
\newcommand{\cn}{\cdot\nabla }
\newcommand{\rot}[1]{\textrm{rot~}#1}	
\renewcommand{\div}[1]{\nabla \cdot #1}	

\numberwithin{equation}{section}

\begin{document}

\centerline{\Large On nonhomogeneous slip boundary conditions}

\smallskip

\centerline{\Large for 2D incompressible exterior fluid flows}

\smallskip

\bigskip

\smallskip

\centerline{Pawe\l ~ Konieczny}

\begin{center}

{Institute of Applied Mathematics and Mechanics}

{Warsaw University}

{ul. Banacha 2, 02-097 Warszawa, Poland}

{E-mail: konieczny@hydra.mimuw.edu.pl}

\end{center}

{\bf Abstract.} 
The paper examines the issue of existence of solutions to the steady Navier-Stokes equations
in an exterior domain in $\mathbb{R}^2$. The system is studied with nonhomogeneous
slip boundary conditions. The main results proves the existence of weak solutions
for arbitrary data. 

\medskip



\section{Introduction}


One of the most difficult problems in the theory of the Navier-Stokes equations are related
to a stationary two dimensional flow in an exterior domain, namely to the problem
\begin{eqnarray}
	v\cdot\nabla v -\nu\Delta v +  \nabla p & = & F \qquad \textrm{in~} \Omega,\label{intr0} \\
	\div v & = & 0 \qquad \textrm{in~} \Omega, \label{intr10}\\
	B(v) & = & 0 \qquad \textrm{on~} \partial\Omega, \label{intr20}\\
	\lim_{|x|\to\infty} v(x) & = & v_\infty,\label{intr30}
\end{eqnarray}
where the sought solution $(v, p)$ is
a velocity vector field and the corresponding pressure,
$\nu$ is a viscous positive constant coefficient,
$F$ -- an exterior force acting on the fluid,
$v_\infty$ -- a prescribed constant vector field
and $B(v)$ stands for a boundary conditions, e.g. Dirichlet boundary conditions, as $v = v_*$ on $\partial\Omega$.
In our case the system will be supplemented with the slip boundary conditions, namely:
\begin{eqnarray}
 	\vec{n}\cdot\mathbb{T}(v, p)\cdot\vec{\tau} + f(v\cdot\vec{\tau}) & = & b\qquad \textrm{on~} \partial\Omega, \label{intr35}\\
 	v\cdot\vec{n} & = & 0 \qquad \textrm{on~} \partial\Omega ,\label{intr37}
\end{eqnarray}
where $f$ is a nonnegative friction coefficient, $\mathbb{T}(v, p)$ is the Cauchy stress tensor, i.e.
$\mathbb{T}(v, p) = \nu\mathbb{D}(v) + p\mathbb{I}$, where
$\mathbb{D}(v) = \{v_{i, j} + v_{j, i}\}_{i, j = 1}^{2}$ is the symmetric
part of the gradient $\nabla v$, and $\mathbb{I}$ is the identity matrix.
Moreover $\vec{n}, \vec{\tau}$ are respectively the normal and tangential
vector to boundary $\partial\Omega$ of an exterior domain $\Omega$,
i.e. $\Omega = \mathbb{R}^2\setminus B$, for a bounded
simply-connected domain $B\subset \mathbb{R}^2$. 

The slip boundary conditions govern the motion of particles at the boundary -- relation
(\ref{intr35}) is just Newton's second law. From the physical point
of view this constraint is more general than the Dirichlet boundary data, since
for $f\to\infty$ and $b\equiv 0$ one can obtain relation $v_{\partial\Omega } = 0$.
The case where $f = 0$ is important for applications, since then the fluid
reacts with surface $\partial\Omega $ as the perfect gas (\cite{ClopMik}, \cite{Mucha1}).

In many modern applications, like the model of motion of blood, polymers and liquid
metals, this type of boundary conditions is widely used (\cite{Fujita}, \cite{Itoh}). 
Our considerations in an exterior 
domain are also important for example in the field of aerodynamics, where problems
with flow past an abstacle is of high interest. 

There are many questions related to this problem, namely: existence of solutions, uniqueness and
asymptotic behaviour. In this paper we are concerned with the first issue for arbitrary data.

We are concerned with weak solutions to the problem (\ref{intr0})-(\ref{intr30}) thus it is natural to require
finite Dirichlet integral
\begin{equation}\label{intr40}
	\int_{\Omega} |\nabla v|^2 dx < \infty.
\end{equation}
Since the power $2$ coincides with the dimension of the domain we are not able to
use standard embedding theorems for $v$ to get some information about the velocity
at infinity and assure that (\ref{intr30}) holds in any sense. 
Many mathematicians brought their attention to this problem. Some partial results
were obtained by Gilbarg and Weinberger (\cite{GW1}, \cite{GW2}), 
like the case with $B(v)$ -- Dirichlet boundary condition with $v_* \equiv 0$ on
$\partial\Omega $. Then the following assertions hold: 
	a) every solution to (\ref{intr0}) - (\ref{intr20}) that satisfies (\ref{intr40}) is
	necessarily bounded;
	b) for every solution to (\ref{intr0}) - (\ref{intr20}) that satisfies (\ref{intr40})
	there exists $\tilde{v}_\infty$ such that
	\begin{equation}\label{intr50}
			\lim_{|x|\to\infty} \int_0^{2\pi} |v(|x|, \theta) - \tilde{v}_\infty|^2 d\theta = 0.
	\end{equation}
In 1988 Amick \cite{Amick1988} published a paper, where he proved that
if the body $B$ is symmetric around the direction of $v_\infty$, and
boundary data $v_*$ is symmetric with respect to the direction of $v_\infty$, then
there exists a symmetric solution $v, p$ such that
\begin{equation}\label{intr60}
	\lim_{|x|\to\infty} v(x) = \tilde{v}_\infty, \textrm{uniformly}.
\end{equation}
These results however give no information about the relation between $v_\infty$ and $\tilde v_\infty$
(see \cite{Galdi} for more detailed information about this problem). 

On the other hand Finn and Smith \cite{FiSm} and Galdi \cite{Galdi1993} showed
that for small values of Reynolds number $\lambda$ and when $v_\infty \neq 0$
there exists at least one solution to the system (\ref{intr0})-(\ref{intr30})
in a proper space. This has been done by applying contraction mapping
technique and proper $L^p$-estimates for the Oseen system in
exterior domain. In our paper we would like to point out
that this technique should also work for our problem considered
with slip boundary conditions. In \cite{2dOseen} we give proper
$L_p$-estimates for the exterior Oseen system with slip boundary conditions.

The main purpose of this paper is to show that the system (\ref{intr0})-(\ref{intr30})
together with (\ref{intr35})-(\ref{intr37}) admits at least one weak solution
for arbitrary data. This result is gatherned in the following theorem:
\begin{Theorem}\label{mainTheorem1}
	Let $\nu > 0$, $f \geq 0$, $F\in (\nabla\dot H^2_0(\Omega))^*$ and
	$b\in H^{-1/2}(\partial\Omega )$. Then for a properly constructed
	vector field $v_0$ there exists at least one
	weak solution (in the sense of Definition \ref{weakDefinition})
	\begin{equation}\label{intr70}
		v = (v_\infty+v_0)+\nabla^\perp\varphi	
	\end{equation}
	to the system (\ref{intr0})-(\ref{intr40}) with
	boundary conditions $B(v)$ as in (\ref{intr35})-(\ref{intr37}),
	for which the following inequalities holds:
	\begin{equation}\label{intr80}
		\|\varphi\|_{\dot{H}^2_0(\Omega)} \leq C = C(\nu, f, b, F) \quad \textrm{and} \quad
		\|\nabla v\|_{L^2(\Omega)}\leq C(\nu, f, b, F).
	\end{equation}
	\textbf{Note:} We used here a simplified notation $\nabla\dot H^2_0(\Omega )$ which stands for:
	\begin{equation}\label{intr90}	
		\nabla\dot H^2_0(\Omega) = \{ \nabla \varphi : \varphi\in \dot H^2_0(\Omega )\} \subset \dot H^2_0(\Omega ).
	\end{equation}
	See also (\ref{H20Definition}).
\end{Theorem}

While dealing with a problem (\ref{intr0})-(\ref{intr30}) one faces a problem of a kernel
of the rot-div operator. In bounded simply connected domains this kernel is trivial
and full information about velocity $v$ can be retrieved from its vorticity $\alpha$. However
in case of unbounded non-simply connected domain some more precise way of finding solution
should be followed. In Section \ref{sub:bounded_domain} we show such results for 
bounded non-simply connected domains, namely we show that the kernel of this
operator, when a solution satisfies the slip boundary conditions, is trivial.

Our construction of $v$ is done with the Galerkin method, what might be useful for
the process of finding the solution numerically. It has one more attribute -- it can be easily
modified to approximate a solution with solutions considered in a sequence of bounded domains,
for which we are able to show that their kernel part of the mentioned operator is trivial.

That is why we may conclude that the cons tructed
vector field is a solution with a trivial kernel part of the rot-div operator. However
we may not exclude the case, that there exists a solution to our problem with
a nontrivial kernel part.

The result from Section \ref{sub:bounded_domain} can be applied to improve results
from \cite{pkpbm}, namely the results from this paper can be extended
(with proper assumptions on the boundary data) to the case of non simply-connected
bounded domain.

One of the classical approaches  to show existence of solutions is to use the Hopf inequality (see \cite{Hopf})
\begin{equation}
	\left\|\frac{u}{\delta}\right\|_{L^2(\Omega )} \leq C\|u\|_{H^1_0(\Omega)}
	\qquad \textrm{for all~}u\in H^1_0(\Omega),
	\label{HopfInequality}
\end{equation}
where $\delta(x) = \mathrm{dist~} (x, \partial\Omega )$ is the distance from the boundary, 
to get a'priori estimates on a solution $u$. However, in case
of slip boundary conditions we are not able to get full information about the velocity
on the boundary $\partial\Omega $ (in particular $u\notin H^1_0(\Omega ))$,
as it is the case for Dirichlet constraint.
Thus different methods are needed to get a'priori estimates.

We introduce a~reformulation of our system in terms
of the vorticity $\alpha = \rot v$ of the fluid. This increase the order of equations
and transforms slip boundary conditions on $v$ into Dirichlet condition on $\alpha$.
This way will be cleared up in the Section \ref{sub:Reformulation}.

Our paper is organized as follows: in Section \ref{sub:bounded_domain} we give some
results about the rot-div operator and a reasoning how to 
improve results from \cite{pkpbm}. In the next section we give a reformulation 
of our problem in terms of rotation
of the fluid, introduce a function space $\dot H^2_0(\Omega )$
and give a definition of a weak solutions to our problem. 
The next section is the main part of our
paper, where we obtain a'priori estimate for our sought vector field. This estimate allows
us to show existence of solutions via the Galerkin method (see: Section \ref{sub:Existence}).

\subsection{The kernel function $\psi$.}\label{sub:bounded_domain} 

\textbf{Note: }We want to emphasize, that the following considerations concern
a bounded domain $\Omega $, while the domain for the main problem (\ref{intr0})-(\ref{intr30})
is an exterior one.

We would like to discuss some details related to
the kernel function $\psi$ for the rot-div operator.
The kernel function $\psi$ appears when one wants to get back
the information about velocity vector field $v$ from its rotation $\alpha$. Namely,
the following system is considered:
\begin{eqnarray}
	\rot v & = & 0 \qquad \textrm{in~} \Omega, \label{bound10}\\
	\div v & = & 0 \qquad \textrm{in~} \Omega, \label{bound20}\\
	v\cdot\vec{n} & = & 0 \qquad \textrm{on~} \partial\Omega. \label{bound30}	
\end{eqnarray}
From (\ref{bound20})-(\ref{bound30}) and from the Poincar\'e Lemma one can take $v$ in the form:
\begin{equation}\label{bound40}
	v = \nabla^\perp \psi,
\end{equation}
for some function $\psi$, which we call the kernel function. From (\ref{bound30})
on can see that
\begin{equation}
	\psi \equiv \textrm{const} \qquad \textrm{on~} \partial\Omega.
\end{equation}
Since $\rot \nabla^\perp\psi = \Delta\psi$ we can figure out that $\psi \equiv 0$
(up to an additive constant) in bounded simply-connected domains.
This was the case in \cite{pkpbm}.
In this section we would like to show that the assumption on simply-connectedness of the domain
in \cite{pkpbm} can be ommited.

Before that we want to point out that the family of solutions to the system (\ref{bound10})-(\ref{bound30}),
in the case of a bounded domain with the first homotopy group $\Pi_1(\Omega) = \mathbb{Z}$, is one dimensional. This
is a simple observation, since all these solutions can be presented in the form
\begin{equation}
	\label{bound50}
	\psi = C_\psi \Psi,
\end{equation}
where $C_\psi$ is a suitable constant and $\Psi$ is the unique solution to the system:
\begin{eqnarray}
 \Delta \Psi & = & 0 \qquad \textrm{in~} \Omega,\\
 \Psi & = & 0 \qquad \textrm{on~} \Gamma_1 ,\\
 \Psi & = & 1 \qquad \textrm{on~} \Gamma_2,
\end{eqnarray}
where $\Gamma_1$ and $\Gamma_2$ are two disjoint connected parts of $\partial\Omega $.

In \cite{2dextlin} it is shown that also for an exterior domain a family
of solutions to a similar problem is also one dimensional. In that case, however,
one needs to use additional property for a solution $v$, namely $v\to 0$ as $|x|\to\infty$.
In the same paper a constant $C_\psi$ is explicity given for a linear modified Oseen problem.

What is interesting is that the slip boundary conditions allow us to calculate that the
kernel part of the solution is equal to zero.
Indeed, in our problem the slip boundary conditions under rot operator transforms into
\begin{equation}\label{bound60}
	\rot v = (2\chi - f/\nu)v\cdot\vec\tau\qquad \textrm{on~} \partial\Omega, 
\end{equation}
thus from (\ref{bound10}) we see that
\begin{equation}\label{bound70}
	(2\chi - f/\nu)\frac{\partial\Psi}{\partial\vec{n}} = 0.
\end{equation}
Since it is impossible that $(2\chi - f/\nu) \equiv 0$ on all $\partial\Omega$ (because of positivity
of $f$ and $\nu$) it has to be
\begin{equation}\label{bound80}
	\frac{\partial\Psi}{\partial\vec{n}} = 0
\end{equation}
at least on an open subset of $\partial\Omega $. This is however impossible since
$\Psi$ is a harmonic function which takes its maximum and minimum on the boundary $\partial\Omega $, and
from the strong Maximum Principle
\begin{equation}
	\frac{\partial\Psi}{\partial\vec{n}} > 0
\end{equation}
at points, where $\Psi$ equals its maximum, which contradicts (\ref{bound80}). The case where $\Psi = \textrm{const}$
in the whole $\Omega $ stays in contradiction with boundary conditions (\ref{bound60})-(\ref{bound70}).

 This conclusion might be used to extend
results from \cite{pkpbm} to the case where the domain is non simply-connected. In this
paper authors are considering simply-connected domains because in this case the kernel of the
rot-div operator is obviously trivial and one may recover full information
about the velocity of the fluid from its rotation.

%


\section{Reformulation}\label{sub:Reformulation}


In this section we give a reformulation of our problem
in terms of the vorticity $\alpha$ of the velocity vector field (see \cite{MuRa}). 

Taking rotation of ($\ref{intr0}$) we get:
\begin{equation}
-\nu\Delta\alpha + v\cn\alpha   =  \rot F, \label{ref10}
\end{equation}
where $\alpha = \rot v = v_{2,1} - v_{1,2}$.
The slip boundary conditions give us full information about the vorticity of the fluid
on the boundary, namely from (\ref{intr40})-(\ref{intr50}) we get a condition on $\alpha$:
\begin{eqnarray}\label{ref40}
\alpha & = & (2\chi - f/\nu)v\cdot \vtau + b\qquad \textrm{on~}\partial\Omega,\\
v\cdot\vn  & = & 0,\label{ref20}\\
v&\to&v_\infty \qquad \textrm{as~} |x|\to+\infty,\label{ref30}
\end{eqnarray}
where $\chi$ is the curvature of the boundary. The exact procedure was considered in \cite{MuRa}.

Next, we take over the information at infinity from $v$. 
Let us introduce an extension vector field $\widetilde{v_0} = v_0 + v_\infty$ satisfying the following conditions:
\begin{eqnarray}
\div{\widetilde{v_0}}  & = & 0\qquad \textrm{in~}\Omega,\label{ref50}\\
\widetilde{v_0} & \equiv & 0\qquad \textrm{on~}\partial\Omega,\label{ref60}\\
\widetilde{v_0} & \to & v_\infty\qquad \textrm{as~}|x|\to\infty.\label{ref70}
\end{eqnarray}
The vector field $v_0$ will be defined later. The construction will fulfil
requirements from Lemma \ref{mainLemma} in order to get a'priori estimates on $\varphi$. 

Having $\widetilde{v_0}$ we rewrite $v$ as
\begin{equation}\label{ref80}
v = \widetilde{v_0} + u,
\end{equation}
where for $u$ we have the following constraints:
\begin{eqnarray}
\div{u}  & = & 0\qquad \textrm{in~}\Omega,\label{ref90}\\
u\cdot\vn  & = & 0\qquad \textrm{on~}\partial\Omega,\label{ref100}\\
u & \to & 0\qquad \textrm{as~}|x|\to\infty.\label{ref110}
\end{eqnarray}
Having (\ref{ref90})-(\ref{ref100}) we use the Poincar\'e lemma 
to present $u$ in the following form:
\begin{equation}\label{ref120}
u = \nabla^\perp \varphi,
\end{equation}
where $\nabla^\perp\varphi = (-\varphi_{,2}, \varphi_{,1})$. 
Since $u\cdot\vn = \frac{\partial\varphi}{\partial\vec{\tau}} = 0$ on $\partial\Omega$ we may take $\varphi \equiv 0$ on $\partial\Omega$. 

For futher calculations let us notice that
\begin{equation}\label{ref130}
\rot\nabla^\perp \varphi  =  \Delta \varphi.
\end{equation}

We may now derive from (\ref{intr0})-(\ref{intr30}) the system of equations for $u$. Recalling that
\begin{equation}\label{ref140}
	\alpha = \rot v = \Delta \varphi +\rot \widetilde{v_0}
\end{equation}
we write:
\begin{equation}
	\label{mainProblem}
	\begin{array}{rcll}
		-\nu\Delta\alpha +~(\widetilde{v_0}+\nabla^\perp\varphi)\cdot\nabla\alpha &=& \rot F & \qquad\mathrm{in~} \Omega, \\
		\alpha - \Delta\varphi &=& \rot{\widetilde{v_0}}& \qquad\mathrm{in~} \Omega,\\
		\varphi & = & 0 & \qquad\mathrm{on~} \partial\Omega,\\
		\alpha -(2\chi - f/\nu)\nabla^\perp\varphi\cdot\vec\tau &=& (2\chi - f/\nu)\widetilde v_0\cdot\vec{\tau} +~b & \qquad\mathrm{on~} \partial\Omega,\\
		\nabla^\perp \varphi & \to & 0 \qquad \mathrm{as~} |x|\to\infty.& 
	\end{array}
\end{equation}
Since we want Dirichlet integral for $v$ to be finite, i.e.
\begin{equation}\label{ref150}
	\int_{\Omega} |\nabla v|^2 ~dx <~\infty,
\end{equation}
we establish a suitable space for the solution $\varphi$.
Let us introduce the following Banach space:
\begin{equation}
	\dot{H}^2_0(\Omega ) = \overline{C_0^\infty(\Omega )}^{\|\nabla^2\cdot\|_{L^2(\Omega )}}
	\label{H20Definition}
\end{equation}
equipped with a~norm
\begin{equation}
	\|\psi\|_{\dot{H}^2_0(\Omega )} = \left( \int_{\Omega} |\nabla^2\psi|^2 ~dx  \right)^{1/2}.
\end{equation}
We now define a weak solution to the problem (\ref{intr0})-(\ref{intr30}).

\begin{Definition}\label{weakDefinition}
	We say that $v$  is a~weak solution
	to problem (\ref{intr0})-(\ref{intr30}) together with
	(\ref{intr40})-(\ref{intr50})
	 iff there exists $\varphi\in \dot{H}^2_0(\Omega )$ 
	such that $v = \widetilde{v_0} + \nabla^\perp\varphi$ and
	the following identity holds for every $\psi\in C^\infty_0(\Omega )$:
\begin{multline}\label{weakphi}
\int_{\Omega} (\nabla^\perp\varphi+\widetilde{v_0})\cdot\nabla\psi (\Delta\varphi+\rot{\widetilde{v_0}})dx+
 \nu \int_{\Omega} (\Delta\varphi+\rot{\widetilde{v_0}}) \Delta\psi~dx+\\
-\nu \int_{\partial\Omega}  (2\chi -f/\nu)(\nabla^\perp\varphi+\widetilde{v_0})
\cdot \tau\frac{\partial \psi}{\partial \vec{n}} d\sigma  =  \int_{\Omega} F\cdot\nabla^\perp\psi ~dx.
\end{multline} 
\end{Definition}
To prove existence of solutions in the sense of Definition \ref{weakDefinition} we
first need to show a'priori estimates on $\varphi$, what will be the case in Section \ref{sub:apriori}.
Before we get into this, we would like to investigate some properties of the kernel function of the rot-div operator,
to explain the properties of the constructed solution.

\subsection{A'priori estimate.}\label{sub:apriori}


To show existence of a solution to our problem we use the standard Galerkin method. This is a construction
of approximate solutions which converge in some sense (strong enough to pass to the limit
in the equation) to the limit vector function.
The construction of this sequence requires usage of a fixed point theorem. Proper converging of this sequence
(or a subsequence) can be assured by showing uniform boundedness (in a proper function space) of all
its elements. In both of these steps a great help is a'priori estimate of solutions to our equation.
That is why we now focus on obtaining it.

We follow the standard approach to get a'priori estimate, i.e. we multiply ($\ref{mainProblem}_1$) by $\varphi$ and
integrate over $\Omega$. Recall that
\begin{equation}\label{apr5}
	\alpha = \Delta \varphi + \rot {\widetilde{v_0}}.	
\end{equation}
The first term $-\nu \int_{\Omega} 	\Delta\alpha\varphi ~dx $ gives us:
\begin{multline}
	-\nu\int_{\Omega} \Delta\alpha\varphi ~dx = -\nu\int_{\Omega} (\Delta\varphi+\rot \widetilde{v_0})\Delta\varphi~dx +\\
	\nu\int_{\partial\Omega} (2\chi -f/\nu) ((\widetilde{v_0}+\nabla^\perp\varphi)\cdot\vec{\tau}+b)\frac{\partial\varphi}{\partial\vec{n}} ~d\sigma .
	\label{firstTerm}
\end{multline}
For the second term it is easily seen that
\begin{equation}\label{apr10}
	\int_{\Omega}(\widetilde{v_0}+\nabla^\perp\varphi)\cdot\nabla\alpha\varphi dx =
	\int_{\Omega} (\widetilde{v_0}+\nabla^\perp\varphi)\cdot\nabla\varphi\alpha ~dx 
\end{equation}
since $\varphi \equiv 0$ on $\partial\Omega $ and $\div {(\widetilde{v_0}+\nabla^\perp\varphi)} = 0$ in $\Omega $.
Finally, since $\nabla^\perp\varphi\cdot\nabla\varphi = 0$ we may write
\begin{equation}\label{apr20}
	\int_{\Omega}(\widetilde{v_0}+\nabla^\perp\varphi)\cdot\nabla\alpha\varphi dx =
\int_{\Omega}\widetilde{v_0}\cdot\nabla\varphi\alpha dx.
\end{equation}
This term causes difficulties in getting a'priori estimates for the solution. The reason why
is that it is of order $2$ with respect to $\varphi$ (see \ref{apr5}), just like $\int_{\Omega} |\Delta\varphi|^2 dx$,
but without any information about its sign. Thus we need to prove an inequality in the form:
\begin{equation}\label{apr30}
	\left|\int_{\Omega} \widetilde{v_0}\cdot\nabla\varphi \Delta\varphi dx\right| \leq \gamma \|\Delta\varphi\|_{L^2(\Omega)}^2,
\end{equation}
for some constant $\gamma$, which should be small enough (in our case $\nu/2$). This is done by a proper
construction of the vector field $\widetilde{v_0}$. 

Since the construction of the vector field $\widetilde{v_0}$ is done in a neighbourhood of $\partial\Omega$
we introduce compactly supported $v_0$ such that
\begin{equation}\label{apr35}
	\widetilde{v_0} = v_0 + v_\infty,
\end{equation}
with proper constraints for $v_0$, which will be precised in the following lemma:
\begin{Lemma}\label{mainLemma}
	For every $\epsilon > 0$ there exists compactly supported $v_0$, which satisfies the following conditions:
	\begin{equation}
		\begin{array}{rcll}
			\div{v_0} &=& 0 &\qquad\mathrm{in~} \Omega, \\
			v_0 &=& -v_\infty &\qquad\mathrm{on~} \partial\Omega,
		\end{array}
	\end{equation}
	and the following inequality holds
	\begin{equation}
		\left| \int_{\Omega} (v_0+v_\infty)\cdot\nabla\varphi\Delta\varphi ~dx \right| \leq
		\epsilon\|\Delta \varphi\|_{L^2(\Omega )}^2
		\label{v0ineq}
	\end{equation}
	for every $\varphi \in \dot{H}^2_0(\Omega)$.
\end{Lemma}

\begin{Proof}{of the Lemma \ref{mainLemma}}
First we transform the term 
\begin{equation}\label{lem1_5}
	\int_{\Omega} \widetilde{v_0}\cn\varphi\Delta\varphi dx,
\end{equation}
using integration by parts, to get a term without $v_\infty$ in it. This is because
the term
\begin{equation}\label{lem1_10}
	\int_{\Omega} v_\infty\cdot\nabla\varphi \Delta\varphi dx = \int_{\Omega} \varphi_{,1}\Delta\varphi dx
\end{equation}
could cause a great difficulty in estimate, since a'priori we do not know whether or not $\varphi_{,1}\in L^2(\Omega)$. 

\textbf{Remark:} in the following calculations we ommit boundary integrals, since $\widetilde{v_0} = 0$ on
$\partial\Omega $, i.e. $v_0^{(1)} = 0$ and $v_0^{(2)}+v_\infty = 0$.
We calculate (\ref{lem1_5})  as follows:
\begin{eqnarray}\label{lem1_20}
\int_{\Omega} \widetilde{v_0}\cdot\nabla\varphi\Delta\varphi  dx& = & 
\int_{\Omega} (v_0+v_\infty)\cdot\nabla\varphi\Delta\varphi dx\\
 & = & \int_{\Omega} v_0^{(1)}\varphi_{,1}\Delta\varphi~
 		+ \int_{\Omega} (v_0^{(2)}+v_\infty)\varphi_{,2}\Delta\varphi~dx\\ =:  I_1 + I_2
\end{eqnarray}
\begin{equation}\label{lem1_30}
I_2  =  \int_{\Omega} (v_0^{(2)}+v_\infty)\varphi_{,2}\varphi_{,11}dx 
	+ \int_{\Omega} (v_0^{(2)}+v_\infty)\varphi_{,2}\varphi_{,22}dx =: I_{21} + I_{22}
\end{equation}
Recalling that $\widetilde{v_0} = 0$ on $\partial\Omega$ we calculate:
\begin{eqnarray}\label{lem1_40}
I_{21}  & = & -\int_{\Omega} (v_0^{(2)}+v_\infty)_{,1}\varphi_{,2}\varphi_{,1}dx -
\int_{\Omega} (v_0^{(2)} + v_\infty)\varphi_{,21}\varphi_{,1}dx \\
 & = & -\int_{\Omega} v_{0,1}^{(2)}\varphi_{,2}\varphi_{,1} dx +
 \frac{1}{2}\int_{\Omega} v_{0,2}^{(2)}\varphi_{,1}^2 dx.
\end{eqnarray}
Similarly $I_{22}$:
\begin{equation}\label{lem1_50}
I_{22}   =  \frac{1}{2}\int_{\Omega} (v_0^{(2)}+v_\infty)(\varphi_{,2}^2)_{,2} dx
 = -\frac{1}{2}\int_{\Omega} v_{0,2}^{(2)}\varphi_{,2}^2 dx.
\end{equation}
For $I_1$ we use the fact that $v_0^{(1)} = 0$ on $\partial\Omega$:
\begin{equation}\label{lem1_60}
I_1 = \int_{\Omega} v_{0}^{(1)}\varphi_{,1}\varphi_{,11} dx+ 
\int_{\Omega} v_0^{(1)}\varphi_{,1}\varphi_{,22} dx=: I_{11} + I_{12}.
\end{equation}
Now we calculate as follows:
\begin{equation}\label{lem1_70}
I_{11} = \frac{1}{2}\int_{\Omega} v_0^{(1)}(\varphi_{,1}^2)_{,1} dx=
-\frac{1}{2}\int_{\Omega} v_{0,1}^{(1)}\varphi_{,1}^2 dx
\end{equation}
and for $I_{12}$:
\begin{equation}\label{lem1_80}
I_{12}   =  -\int_{\Omega} v_{0, 2}^{(1)}\varphi_{,1}\varphi_{,2}dx
-\int_{\Omega} v_0^{(1)}\varphi_{,12}\varphi_{,2} dx
 = -\int_{\Omega} v_{0, 2}^{(1)}\varphi_{,1}\varphi_{,2}dx
  + \frac{1}{2}\int_{\Omega} v_{0, 1}^{(1)}\varphi_{,2}^2dx.
\end{equation}
Finally, summing up all above calculations we get:
\begin{eqnarray}
I  & = & -\frac{1}{2}\int_{\Omega} v_{0,1}^{(1)}\varphi_{,1}^2dx
		-\int_{\Omega} v_{0, 2}^{(1)}\varphi_{,1}\varphi_{,2}dx
		+ \frac{1}{2}\int_{\Omega} v_{0, 1}^{(1)}\varphi_{,2}^2dx \\
	&&  -\int_{\Omega} v_{0, 1}^{(2)}\varphi_{,1}\varphi_{,2}dx
		+\frac{1}{2}\int_{\Omega} v_{0, 2}^{(2)}\varphi_{,1}^2dx
		-\frac{1}{2}\int_{\Omega} v_{0, 2}^{(2)}\varphi_{,2}^2dx\\
	 & = & -\frac{1}{2}\int_{\Omega} \varphi_{,1}^2(v_{0, 1}^{(1)} - v_{0, 2}^{(2)})dx
	 	+ \frac{1}{2}\varphi_{,2}^2(v_{0, 1}^{(1)} - v_{0, 2}^{(2)})dx\\
	 && -\int_{\Omega} \varphi_{,1}\varphi_{,2}(v_{0, 2}^{(1)} + v_{0, 1}^{(2)})dx.
\end{eqnarray}
Now, since $\div v_0 = 0$ we may write:
\begin{eqnarray}\label{lem1_90}
I  & = &  -\int_{\Omega} \varphi_{,1}^2 v_{0, 1}^{(1)}dx - \int_{\Omega} \varphi_{,2}^2 v_{0, 2}^{(2)}dx - \int_{\Omega} \varphi_{,1}\varphi_{,2}(v_{0, 2}^{(1)} + v_{0, 1}^{(2)})dx\\
  & = & -\int_{\Omega} \nabla\varphi\cdot\nabla v_0\cdot\nabla\varphi dx
\end{eqnarray}

In this form we see, that there is no term with $v_\infty$, but its structure
does not allow us to go into more subtle analysis of its behavior near the boundary of the domain.
That is why we transform it into more appriopriate term. This is not straightforward
since simple calculation by parts would lead us to the point we have started with. This is because
there is still information about $v_\infty$ in this term -- it occurs in $v_0$ on the boundary.
Thus an auxilary vector field is needed to take away  $v_\infty$ from the boundary. We proceed
as follows:
\begin{equation}\label{lem1_100}
I = -\int_{\Omega} \nabla\varphi\cdot\nabla v_0\cdot\nabla\varphi =
-\int_{\Omega} \nabla\varphi\cdot\nabla(v_0+V_\epsilon)\cdot\nabla\varphi +
\int_{\Omega} \nabla\varphi\cdot\nabla V_\epsilon\cdot\nabla\varphi,
\end{equation}
where $V_\epsilon$ is contructed as follows:
let us introduce a vector field $V$ in $(t_1, t_2)$ coordinates (see Appendix):
\begin{equation}\label{lem1_110}
V(p(t_1, t_2)) := {\frac{v_{\infty}}{2}[1+cos((\pi/\zeta) t_2)] \choose 0}
\end{equation}
for which the following conditions are valid:
\begin{equation}\label{lem1_120}
V(p(t_1, 0))  = {v_\infty \choose 0}, \qquad
V(p(t_1, \zeta)) =  {0 \choose 0},
\end{equation}
where $\zeta = \zeta(\Omega)$ is a constant from the construction of the mapping $p(t_1, t_2)$. 

The similar conditions are fulfilled by the vector field
\begin{equation}\label{lem1_130}
V_\epsilon(p(t_1, t_2)) := 
\left\{
\begin{array}{ll}
V(p(t_1, t_2/\epsilon)) & \textrm{for~}t_2\leq \zeta\epsilon \\
0 & \textrm{for~}\zeta\epsilon \leq t_2 \leq \zeta.
\end{array}\right.
\end{equation}
From the construction of $V_\epsilon$ it is easily seen that
\begin{equation}\label{nablaVepsineq}
\|\nabla V_\epsilon\|_{L^2(\Omega)} \leq C(\Omega, V)\frac{1}{\epsilon^{1/2}}.
\end{equation}
Indeed, from the definition (\ref{lem1_130}) we calculate:
\begin{eqnarray}
 \int_{\Omega} |\nabla V_\epsilon|^2 &\leq &
 		C(p) \int_0^L \int_0^{\zeta\epsilon} |\nabla_t(V(p(t_1, t_2/\epsilon)))|^2 dt_2dt_1\\
 		& \leq & C(p)\int_0^L \int_0^{\zeta\epsilon} |\nabla_t V(p(t_1, t_2/\epsilon))|^2\frac{1}{\epsilon^2} dt_2 dt_1\\
 		& = & C(p)\int_0^L \int_0^{\zeta}|\nabla V|^2 \frac{1}{\epsilon} dt_2' dt_1 \leq \frac{C(p)}{\epsilon}\|\nabla V\|_{L^2(\Omega)}^2.
\end{eqnarray}
We may now estimate the integral $I$:
\begin{equation}\label{lem1_140}
I   =  -\int_{\Omega} \nabla\varphi\cdot\nabla(v_0+V_\epsilon)\cdot\nabla\varphi dx
 + \int_{\Omega} \nabla\varphi\cdot\nabla V_\epsilon\cdot\nabla\varphi dx=: I_1 + I_2.
\end{equation}
Since $v_0+V_\epsilon = 0$ on $\partial\Omega$ we may integrate $I_1$ by parts and get:
\begin{equation}\label{lem1_150}
I_1 = \int_{\Omega} \nabla\varphi (v_0+V_\epsilon)\Delta\varphi dx+ \frac{1}{2}\int_{\Omega} ((v_0+V_\epsilon)\cdot\nabla)(|\nabla\varphi|^2) dx = I_{11} + I_{12}
\end{equation}
Now, from the Stokes theorem and since $\div v_0 = 0$ in $\Omega$
\begin{equation}\label{lem1_160}
I_{12} = -\frac{1}{2}\int_{\Omega} (\div V_\epsilon)(|\nabla\varphi|^2) dx.
\end{equation}
Gathering all these calculations we get:
\begin{multline}\label{lem1_170}
I  =  \int_{\Omega} \nabla\varphi v_0\Delta\varphi dx+
\int_{\Omega} \nabla\varphi V_\epsilon\Delta\varphi dx-
\frac{1}{2}\int_{\Omega} (\div V_\epsilon)(|\nabla\varphi|^2) dx+\\
+\int_{\Omega} \nabla\varphi\cdot\nabla V_\epsilon\cdot\nabla\varphi dx
=: J_1 + J_2 + J_3 + J_4.
\end{multline}
Before we estimate these integrals let us introduce the following notation:
\begin{equation}\label{lem1_180}
\Omega_\epsilon = \{x\in \Omega : \textrm{dist}(x, \partial\Omega) < \epsilon\zeta\}.
\end{equation}
Integrals $J_3$ and $J_4$ are similar and we estimate them first. Since $\textrm{supp~} V_\epsilon\subset \Omega_\epsilon$:
\begin{eqnarray}\label{I3I4ineq}
J_3 + J_4 \leq C \left(\int_{\Omega_\epsilon} |\nabla V_\epsilon|^2\right)^{1/2}
\left(\int_{\Omega_\epsilon} |\nabla\varphi|^4\right)^{1/2}
\end{eqnarray}
We use the following interpolation inequality:
\begin{equation}\label{interpineq}
c\|\nabla\varphi\|_{L^4(\Omega_\epsilon)} \leq \|\varphi\|^{1/4}_{L^2(\Omega_\epsilon)}\|\nabla^2\varphi\|^{3/4}_{L^2(\Omega_\epsilon)}.
\end{equation}
Since $\varphi\equiv 0$ on $\partial\Omega$ and $\varphi\in H^2(\Omega_\epsilon)$ we use embedding theorem 
\begin{equation}\label{lem1_190}
H^2(\Omega_\epsilon)\subset C^{\alpha}(\Omega_\epsilon),\qquad \alpha < 1
\end{equation}
to conclude that
\begin{equation}\label{lem1_200}
\|\varphi\|_{L^2(\Omega_\epsilon)} \leq C(\Omega_\epsilon)\|\nabla^2\varphi\|_{L^2(\Omega_\epsilon)}\cdot\epsilon^{\frac{1+2\alpha}{2}}.
\end{equation}
Indeed:
\begin{eqnarray}\label{lem1_210}
\|\varphi\|^2_{L^2(\Omega_\epsilon)}  & \leq & 
\int_{0}^L \int_{0}^{\zeta\epsilon} |\varphi(t_1, t_2)|^2 |Jp|dt_1 dt_2\\
 & \leq & C(\Omega)\int_{0}^L \int_{0}^{\zeta\epsilon} t_2^{2\alpha}
 \|\nabla^2\varphi\|^2_{L^2(\Omega_\epsilon)}dt_1 dt_2\\
  & \leq & C(\Omega) \|\nabla^2\varphi\|^2_{L^2(\Omega_\epsilon)}\epsilon^{1+2\alpha}.
\end{eqnarray}
Inserting this inequality to (\ref{interpineq}) we get:
\begin{equation}\label{phiL4ineq}
\|\nabla\varphi\|_{L^4(\Omega_\epsilon)}^2\leq C(\Omega)\|\nabla^2\varphi\|_{L_2(\Omega_\epsilon)}^2\cdot\epsilon^{\frac{1+2\alpha}{4}}.
\end{equation}
Now from (\ref{nablaVepsineq}), (\ref{I3I4ineq}) and (\ref{phiL4ineq}) we conclude:
\begin{equation}\label{lem1_220}
J_3 + J_4 \leq C(\Omega, V) \epsilon^{\frac{2\alpha - 1}{4}}\|\nabla^2\varphi\|_{L_2(\Omega)}^2.
\end{equation}
Since $0 < \alpha < 1$ (in particular $\alpha$ can be taken $\alpha > 1/2$) we may choose $\epsilon$ small enough to get
\begin{equation}\label{lem1_230}
J_3 + J_4 \leq \frac{1}{8}\|\nabla^2\varphi\|_{L_2(\Omega)}^2.
\end{equation}
The estimate of the integral $J_2$ is similar:
\begin{equation}\label{lem1_240}
J_2   =  \int_{\Omega} \nabla\varphi V_\epsilon\Delta\varphi\\ 
 \leq  \left(\int_{\Omega_\epsilon} |V_\epsilon|^4~dx\right)^{1/4}
 \left(\int_{\Omega_\epsilon} |\nabla\varphi|^4~dx\right)^{1/4}
 \left(\int_{\Omega} |\nabla^2\varphi|^2\right)^{1/2}
\end{equation}
and since 
\begin{equation}\label{lem1_250}
\left(\int_{\Omega_\epsilon} |V_\epsilon|^4~dx\right)^{1/4}\leq
\left(\int_{\Omega} |V|^4~dx\right)^{1/4} = C(V)\leq \infty
\end{equation}
we may combine (\ref{phiL4ineq}) with (\ref{lem1_250}) and (\ref{lem1_240}), to get, for $\epsilon$ small enough, desired estimate:
\begin{equation}\label{lem1_260}
	J_2 \leq \frac{1}{8} \|\nabla^2\varphi\|^2_{L^2(\Omega)}
\end{equation}
For integral $J_1$ we refer the Reader to Lemma 2.1. from \cite{pkpbm}, which states that
for every $\epsilon > 0$ the vector field $v_0$ can be constructed in such a way
that the following inequality holds:
\begin{equation}\label{lem1_270}
	\left| \int_{\Omega} (v_0\cdot\nabla\varphi)^2 dx\right| \leq \epsilon \|\Delta\varphi\|_{L^2(\Omega )}^2
	\qquad \textrm{for all~}\varphi\in\dot{H}^2_0(\Omega ).
\end{equation}
Thus integral $J_1$ can be estimated using Schwarz inequality and above lemma:
\begin{equation}\label{lem1_280}
	\left|\int_{\Omega} \nabla\varphi v_0\Delta\varphi dx\right| \leq \frac{1}{4} \|\nabla^2\varphi\|_{L^2(\Omega)}^2
\end{equation}

This completes the proof of Lemma \ref{mainLemma}.
\end{Proof}

Above inequalities allow us to get a'priori estimates. Namely $\varphi$ fulfills the following
identity
\begin{multline}
	-\nu\int_{\Omega} |\Delta\varphi|^2 ~dx 
	+\nu \int_{\partial\Omega} ((2\chi - f/\nu)((\widetilde{v_0}+\nabla^\perp\varphi)\cdot \vec{\tau}) + b)
	\frac{\partial\varphi}{\partial\vec{n}}d\sigma +\\
	+\int_{\Omega} \widetilde{v_0}\cdot\nabla\varphi(\rot\widetilde{v_0}+\Delta\varphi) ~dx 
	= \nu \int_{\Omega}  (\rot \widetilde{v_0})\Delta\varphi~dx + \int_{\Omega} F\cdot\nabla^\perp\varphi ~dx .
	\label{mainEstimateEquation}
\end{multline}
We see, that there is still a terms of order two with respect to $\varphi$, for which one cannot verify its sign,
namely
\begin{equation}\label{lem1_290}
	\int_{\partial\Omega } 2\chi (\nabla^\perp\varphi)\cdot\vec{\tau}\frac{\partial\varphi}{\partial\vec{n}} d\sigma =
	\int_{\partial\Omega } 2\chi \left(\frac{\partial\varphi}{\partial\vec{n}}\right)^2.
\end{equation}
We deal with this problem using the following identity (see \cite{pkpbm}) for $v = \widetilde{v_0}+u$:
\begin{equation}
	\int_{\Omega} \alpha^2 ~dx - \int_{\partial\Omega} \alpha(v\cdot \vec{\tau}) ~d\sigma =
	\int_{\Omega} \mathbb{D}^2(v) ~dx + \int_{\partial\Omega} \left( (v\cdot \vec{\tau})^2 f - b(v\cdot\vec{\tau}) \right) ~d\sigma,
	\label{alphaEquation}
\end{equation}
which comes from the well known identity for $v\in \dot{H}^1(\Omega)$ with $\div v = 0$ in $\Omega$ (see \cite{Solonnikov}):
\begin{equation}\label{lem1_292}
	\int_{\Omega} \alpha^2 dx = \int_{\Omega} |\mathbb{D}(v)|^2 dx + \int_{\partial\Omega }2\chi(v\cdot\vec{\tau})^2.
\end{equation}
Since	
\begin{multline}
	-\nu\int_{\Omega} |\Delta\varphi|^2 ~dx +
	\nu\int_{\partial\Omega} ((2\chi-f/\nu)((\widetilde{v_0}+\nabla^\perp\varphi)\cdot\vec{\tau})+b)\left( \frac{\partial\varphi}{\partial \vec{n}} \right)~d\sigma - \int_{\Omega} \Delta\varphi\rot{\widetilde{v_0}} ~dx=\\
	-\nu\int_{\Omega} \alpha^2 ~dx +\nu\int_{\partial\Omega} \alpha(v\cdot\vec{\tau}) ~d\sigma 
	+\int_{\Omega} \Delta\varphi\rot{\widetilde{v_0}} ~dx +\int_{\Omega} (\rot\widetilde{v_0})^2 ~dx 
	 -\nu\int_{\partial\Omega} \alpha(v_0\cdot\vec{\tau}) ~d\sigma,
\end{multline}
we may derive from (\ref{mainEstimateEquation}) using (\ref{alphaEquation}) the following identity:
\begin{eqnarray}\label{lem1_295}
		-\nu \int_{\Omega} \mathbb{D}^2(\nabla^\perp\varphi + \widetilde{v_0}) ~dx+&&\nonumber\\ 
+\int_{\Omega} \Delta\varphi\rot{\widetilde{v_0}} ~dx +\int_{\Omega} (\rot\widetilde{v_0})^2 ~dx
		 &&\nonumber\\
	-\nu\int_{\partial\Omega} ( (\nabla^\perp\varphi + \widetilde{v_0})\cdot\vec{\tau})^2f 
		- b( (\nabla^\perp\varphi+\widetilde{v_0})\cdot\vec{\tau})~d\sigma-&&\\ 
	 -\nu\int_{\partial\Omega} \alpha(v_0\cdot\vec{\tau}) ~d\sigma
	 +\int_{\Omega} \widetilde{v_0}\cdot\nabla\varphi(\rot\widetilde{v_0}+\Delta\varphi) ~dx
	 &=& \int_{\Omega} F\cdot\nabla^\perp\varphi ~dx \nonumber.
\end{eqnarray}
To get a'priori estimate from (\ref{lem1_295}) we need to use the Korn's inequality (see Lemma \ref{korn} in Appendix):
\begin{equation}\label{lem1_300}
	\int_{\Omega} \mathbb{D}^2(\nabla^\perp\varphi) dx \geq K \int_{\Omega} |\nabla^2 \varphi|^2 dx,
\end{equation}
where $K$ is a constant dependent only on $\Omega $, 
which allows us to get $\int_{\Omega} |\nabla^2\varphi|^2dx$ in the estimate, namely:
\begin{equation}\label{lem1_310}
 	-\nu \int_{\Omega} \mathbb{D}^2(\nabla^\perp\varphi) dx 
 	-\nu\int_{\partial\Omega} ( (\nabla^\perp\varphi + \widetilde{v_0})\cdot\vec{\tau})^2f 
 	\leq -K\nu\int_{\Omega} |\nabla^2\varphi|^2 dx.
 \end{equation} 
 Here we also used the fact that $f\geq 0$.  Combining (\ref{lem1_295}) with (\ref{lem1_310}) we 
are able to get an estimate of the form:
\begin{equation}\label{lem1_320}
	\int_{\Omega} |\nabla^2\varphi|^2 dx \leq C(\textrm{DATA}) \left(\int_{\Omega} |\nabla^2\varphi|^2\right)^{1/2}
		+ \frac{1}{K\nu}\left|\int_{\Omega} \widetilde{v_0}\cdot\nabla\varphi\Delta\varphi dx\right|.
\end{equation}
This is because $\left|\int_{\Omega} \widetilde{v_0}\cdot\nabla\varphi\Delta\varphi dx\right|$ is the
only term of order $2$ with respect to $\varphi$, and all other terms are of order $1$ can be estimated by
$C(\textrm{DATA})\left(\int_{\Omega} |\nabla^2\varphi|^2\right)^{1/2}$ (see the Remark below).

We now use Lemma \ref{mainLemma} with $\epsilon = K\nu/2$ to estimate remaining term
of order $2$ and get the following inequality:
\begin{equation}\label{lem1_330}
	\|\nabla^2\varphi\|_{L^2(\Omega )}^2 \leq C(\mathrm{DATA})\|\nabla\varphi\|_{L^2(\Omega )},
\end{equation}
where in $C(\mathrm{DATA})$ one includes all constants dependent on $\Omega$, $F$, $\nu$, etc.
This inequality gives us of course a'priori estimate on $\|\nabla^2\varphi\|_{L^2(\Omega )}$:
\begin{equation}\label{lem1_340}
	\|\nabla^2\varphi\|_{L^2(\Omega) } \leq C(\mathrm{DATA}).
\end{equation}

\textbf{Remark:} It is not hard to estimate terms in (\ref{lem1_300}), where $u$ (i.e. $\nabla^\perp\varphi$) appears in a~form different than
$\nabla^2\varphi$ -- one must recall that $\varphi \equiv 0$ on $\partial\Omega $, which gives
us, together with $\varphi\in H^2_\mathrm{loc} (\Omega )$, the fact, that all local estimates (the only
needed) can be obtained using $\|\nabla^2\varphi\|_{L^2(\Omega )}$.

\section{Existence.}\label{sub:Existence}


In this section we use standard Galerkin method to prove the existence of a solution to 
\begin{eqnarray}\label{rotns}
-\nu\Delta\alpha + v\cn\alpha  & = & \rot F, \\
v\cdot\vn  & = & 0,\\
\textrm{rot~}v  & = & \alpha,\qquad \textrm{in~}\Omega\\
\div v & = & 0,\qquad \textrm{in~}\Omega\\
\alpha  & = &  (2\chi - f/\nu)v\cdot\vtau +b\qquad \textrm{on~}\partial\Omega
\end{eqnarray}
in the sense of distributions, i.e. in the sense of Definition \ref{weakDefinition}. This
means that we prove the existence of a function $\varphi\in \dot{H}^2_0(\Omega)$, which
satisfies (\ref{weakphi}) for all $\psi \in C^\infty(\Omega )$ with compact support
in $\Omega $. As was mentioned before, we seek for a solution in the form
\begin{equation}\label{exist10}
	v = \widetilde{v_0}+\nabla^\perp\varphi.
\end{equation}

Since $\dot H^2_0(\Omega)$ is Hilbertian and separable we take a base 
of compactly supported functions $\{w_i\}_{i = 1}^\infty$:
\begin{equation}
\dot H^2_0(\Omega) = \overline{\{w_1, w_2, \ldots\}}^{\|\cdot\|_{\dot H^2_0(\Omega)}}.
\end{equation}
Next we introduce a finite dimensional subspace $V^N(\Omega) \subset \dot H^2_0(\Omega)$:
\begin{equation}
V^N(\Omega) = span\{w_1, \ldots, w_N\}.
\end{equation}
We additionally assume that $(w_i, w_j)_{V^N} = \delta_{ij}$, where $(\cdot, \cdot)_{V^N}$ is the inner product
in $V^N$, which comes from the inner product in $\dot H^2_0(\Omega )$. 
We search for an approximation $\varphi^N$ of the function $\varphi$ in the form:
\begin{equation}
\varphi^N(x) = \sum_{j=0}^N c_j^N w_j \in V^N.
\end{equation}
To find coefficients $c_j^N$ we solve the following system:
\begin{multline}
\int_{\Omega} (\nabla^\perp\varphi^N+\widetilde{v_0})\cdot\nabla w_i (\Delta\varphi^N+\rot{\widetilde{v_0}})dx+
 \nu \int_{\Omega} (\Delta\varphi^N+\rot{\widetilde{v_0}}) \Delta w_i~dx+\\
-\nu \int_{\partial\Omega}  \left((2\chi -f/\nu)(\nabla^\perp\varphi^N+\widetilde{v_0})
\cdot \vec\tau+b\right)\frac{\partial  w_i}{\partial \vec{n}} d\sigma  =  \int_{\Omega} F\cdot\nabla^\perp w_i ~dx
\end{multline}
for $i = 1, \ldots, N$.

Let us introduce a mapping $P : V^N \to V^N$ as follows:
\begin{eqnarray*}
P(\varphi^N) & = & \sum_{i = 1}^N\left ( \int_{\Omega} (\nabla^\perp\varphi^N+\widetilde{v_0})\cdot\nabla w_i (\Delta\varphi^N+\rot{\widetilde{v_0}})dx \right . \\
&&  +\nu \int_{\Omega} (\Delta\varphi^N+\rot{\widetilde{v_0}}) \Delta w_i~dx\\
&&-\nu \int_{\partial\Omega}  \left((2\chi -f/\nu)(\nabla^\perp\varphi^N+\widetilde{v_0})
\cdot\vec\tau+b\right)\frac{\partial  w_i}{\partial \vec{n}} d\sigma\\
&&  \left .- \int_{\Omega} F\cdot\nabla^\perp w_i ~dx\right )\cdot w_i.
\end{eqnarray*}
From the definition of the mapping $P$ we easily calculate that:
\begin{eqnarray*}
\left(P(\varphi^N),\varphi^N\right)_{V^N} & = & \int_{\Omega} (\nabla^\perp\varphi^N+\widetilde{v_0})\cdot\nabla\varphi^N (\Delta\varphi^N+\rot{\widetilde{v_0}})dx \\
&&  +\nu \int_{\Omega} (\Delta\varphi^N+\rot{\widetilde{v_0}}) \Delta \varphi^N~dx\\
&&-\nu \int_{\partial\Omega} \left( (2\chi -f/\nu)(\nabla^\perp\varphi^N+\widetilde{v_0})
\cdot\vec\tau+b\right)\frac{\partial  \varphi^N}{\partial \vec{n}} d\sigma\\
&&  - \int_{\Omega} F\cdot\nabla^\perp \varphi^N ~dx
\end{eqnarray*}
To get proper estimate for the term $(P(\varphi^N), \varphi^N)_{V^N}$ first we must use
the following identity (see (\ref{alphaEquation}) or (\ref{lem1_292})):
\begin{multline}\label{alphaEq}
\int_{\Omega} \alpha^2~dx - \int_{\partial\Omega} \alpha(v\cdot\tau)d \sigma=
\int_{\Omega} {\mathbb{D}}^2(v)~dx +
\int_{\partial\Omega} ((v\cdot\tau)^2f - b(v\cdot\vec\tau))d \sigma.
\end{multline}
Hence, since $\nabla^\perp\varphi^N\cdot\nabla\varphi^N = 0$:
\begin{multline}
\left(P(\varphi^N),\varphi^N\right)_{V^N} = \int_{\Omega} \widetilde{v_0}\cdot\nabla\varphi^N (\Delta\varphi^N+\rot{\widetilde{v_0}})dx -\nu \int_{\Omega} \alpha\rot{\widetilde{v_0}}\\
\nu \left(\int_{\Omega} {\bf D}^2(v)dx + \int_{\partial\Omega} ((v\cdot\vec{\tau})^2f - b(v\cdot\vec\tau))\right)d\sigma
- \int_{\Omega} F\cdot\nabla^\perp \varphi^N ~dx
\end{multline}
In this form it is easily seen, that one can obtain the following estimate
\begin{equation}
	\left( P(\varphi^N), \varphi^N \right)_{V^N} > 0\qquad \mathrm{for~} \|\varphi^N\|_{V^N} > k	
\end{equation}
for some constant $k > 0$. One must just repeat the reasoning from a'priori estimates.

Such condition gives us (see \cite{Konieczny}) existence of $\varphi_N$ such that $\|\varphi^N\|_{V^N} \leq k$ and moreover $P(\varphi^N) = 0$, which solves our problem for coefficients $c_j^N$.
Thus we get sequence of approximating solutions $\varphi^N \in V^N$ such that
$\|\varphi^N\|_{V^N} \leq k$ for some constant $k > 0$ independent of $N$.

\textbf{Passing to the limit.} Since we have uniform bound on $\|\varphi^N\|_{\dot H^2_0(\Omega)}$,
i.e. $\|\varphi^N\|_{\dot H^2_0(\Omega)} < k$ (we can take a
subsequence which is weakly convergent to some limit. However, for the sake of passing to the limit in
nonlinear terms of (\ref{weakphi}) we must use diagonal technique.

Let us denote $\Omega_R = \Omega \cap B_R$. In bounded domain $\Omega_R$ we have
$\varphi^N \in H^2(\Omega_R)$, since $\varphi^N, \nabla\varphi^N\in L^2_\textrm{loc}(\Omega)$.
Moreover, since $\varphi^N \equiv 0$ on $\partial\Omega$ we have
\begin{equation}
\|\varphi^N\|_{H^2(\Omega_R)} \leq C(R)\|\varphi^N\|_{\dot H^2(\Omega)},
\end{equation}
hence we may choose a subsequence $\varphi^{N'}$, which we futher denote
for simplicity as $\varphi^N$, which is convergent on $\Omega_R$ to $\varphi$
in the following sence:
\begin{eqnarray}
\nabla^2\varphi^N & \to & \nabla^2\varphi \quad \textrm{weakly~} L^2 \textrm{~on~} \Omega_R,\\
\nabla\varphi^N & \to & \nabla\varphi \quad \textrm{strongly~} L^2 \textrm{~on~} \Omega_R,\\
\varphi^N & \to & \varphi \quad \textrm{strongly~} L^2 \textrm{~on~} \Omega_R,\\
\varphi^N & \to & \varphi \quad \textrm{strongly~} L^2 \textrm{~on~} \partial\Omega.
\end{eqnarray}
We repeat this treatment for $R \to \infty$ to get subsequence $\varphi^N$, which is
convergent in above sense on all bounded domains. Since a test function in $\ref{weakphi}$
has compact support we may pass to the limit. Thus $\varphi$ is
a solution in the sense of Definition \ref{weakDefinition}.

\section{Appendix}


{\large \textbf{$(t_1, t_2)$-coordinates}}\label{app10}\bigskip

We introduce \textbf{$(t_1, t_2)$-coordinates} as follows.
Let $s: [0, L] \to \mathbb{R}^2$ be a normal parameterization of 
 boundary $\partial\Omega$, i.e.
\begin{eqnarray}\label{sdef}
s([0, L])   =  \partial\Omega,\quad
s(0) = s(L)   =  x_0 \in \partial\Omega, \quad {\rm and} \quad
|s'(t)|   =  1
\end{eqnarray}
for a fixed point $x_0$ and $L$ -- the length of $\partial \Omega$. 
Next we introduce the following mapping
$p : [0, L]\times [0, \zeta] \to \mathbb{R}^2$ such that
\begin{equation}\label{pdef}
p(t_1, t_2) = s(t_1) + t_2 \vec{n}(s(t_1)),
\end{equation}
where $\vec{n}$ is the inner normal vector to boundary $\partial\Omega$.
If $\zeta$ is small enough (comparing to curvature $\chi$ of 
 boundary $\partial\Omega$), then the map is one-to-one and $p\in C^1$. 
 Moreover
\begin{equation}\label{dist}
dist(p(t_1, t_2), \partial\Omega) = t_2.
\end{equation}
Using the definition  we compute the gradient of map $p$ as follows
\begin{equation}\label{pgrad}
p_{,{t_1}}   =  (1+t_2\chi)\vec{\tau}(s(t_1)),\qquad
p_{,{t_2}}   =  \vec{n}(s(t_1)).
\end{equation}
Then we see that \begin{equation}\label{perp}
p_{,1} \perp p_{,2} \quad
{\rm and} \quad
(\nabla p)^{-1}=(\frac{1}{1+t_2\chi}\vec{\tau},\vec{n})^T.
\end{equation}
By coordinates $(t_1, t_2)$ we denote coordinates obtained using mapping $p$.

\bigskip

\noindent{\large \textbf{Korn's inequality}}

\bigskip

\begin{Lemma}\label{korn}
	For an exterior domain $\Omega \subset \mathbb{R}^2$, which is not spherically symmetric,
	there exists a~constant $K > 0$, dependent on the domain $\Omega$ such that
	the following inequality
	\begin{equation}
		\int_{\Omega} \mathbb{D}^2(u) ~dx \geq K\int_{\Omega} |\nabla u|^2 ~dx 
		\label{kornIneq}
	\end{equation}
	holds for every $u \in \overline{C_0^\infty}^{\|\nabla\cdot\|_{L^2(\Omega )}}$ satisfying
	\begin{equation}
		\label{kornReq}
		\div u = 0\quad \mathrm{in~} \Omega,\qquad u\cdot\vec{n} = 0 \quad\mathrm{on~} \partial\Omega.
	\end{equation}
\end{Lemma}

See \cite{Solonnikov} for the proof of this lemma.

\textsl{Acknowledgement.} 
The author would like to thank Piotr Mucha for useful discussions during preparation of this paper.
The paper has been supported by Polish grant No. N201 035 32/2271.


{\footnotesize
\begin{thebibliography}{99}


\bibitem[1]{Amick1988} Amick, C.J.: On Leray's Problem of Steady Navier-Stokes
Flow Past a Body in the Plane, Acta Math., 161 (1988), 71--130.

\bibitem[2]{BorPil} Borchers, W., Pileckas, K., Note on the Flux Problem for Stationary
Incompressible Navier-Stokes Equations in Domains with Multiply Connected
Boundary, Acta Appl. Math. 37 (1994), 21--30.

\bibitem[3]{ClopMik} Clopeau, T., Mikeli\'c, A., Robert, R., On the vanishing 
viscosity limit for the 2D incompressible Navier-Stokes equations 
with the friction type boundary conditions. Nonlinearity 11 (1998), no. 6,
 1625--1636.

\bibitem[4]{Farwig} Farwig, R.: Stationary solutions of compressible Navier-Stokes
equations with slip boundary conditions, Comm. PDE 14, (1989) 1579--1606

\bibitem[5]{FiSm} Finn, R., Smith, D.R.; On the Stationary Solution of the
Navier-Stokes Equations in Two Dimensions, Arch. Rational Mech. Anal. 25 (1967)
26--39.

\bibitem[6]{Fujita} Fujita, H., Remarks on the Stokes flow under slip and leak boundary 
conditions of friction type.  Topics in mathematical fluid mechanics,  73--94,
 Quad. Mat., 10,  2002.

\bibitem[7]{Galdi1993} Galdi, G.P.: Existence and Uniqueness at Low Reynolds
Number of Stationary Plane Flow of a Viscous Fluid in Exterior Domains.
Recent Developments in Theoretical Fluid Mechanics, Galdi, G.P., and
Necas, J., Eds., Pitman Research Notes in Mathematics Series,
Longman Scientific and Technical, Vol. 291 (1993), 1--33.

\bibitem[8]{Galdi} Galdi, G.P.: An Introduction to the Mathematical Theory of the
Navier-Stokes Equations, Springer Tracts in Natural Philosophy, 1994.

\bibitem[9]{GW1} Gilbarg, D.; Weiberger, H.F.: Asymptotic Properties of Leray's
Solution of the Stationary Two-Dimensional Navier-Stokes Equations.
Russian Math. Surveys, 29 (1974), 109--123.

\bibitem[10]{GW2} Gilbarg, D.; Weiberger, H.F.: Asymptotic Properties of Steady Plane
Solutions of the Navier-Stokes Equations with Bounded Dirichlet Integral.
Ann. Scuola Norm. Sup. Pisa, (4), 5 (1978), 381--404.

\bibitem[11]{Hopf} Hopf, E., Ein allgemeiner Endlichkeitssatz der Hydrodynamik, 
Math. Ann. 117 (1941), 764--775.

\bibitem[12]{Itoh} Itoh, S.; Tanaka N.; Tani A.: The initial value problem
for the Navier-Stokes equations with general slip boundary
condition, Adv. Math. Sci. Appl. 4, (1994) 51--69

\bibitem[13]{2dextlin} Konieczny, P., Linear flow problems in 2D exterior
domain for 2D incompressible fluid flows, Banach Center Publ., to appear.

\bibitem[14]{pkpbm} Konieczny, P.; Mucha, P. B., On nonhomogeneous slip boundary conditions
for 2D incompressible fluid flows, Internat. J. Engrg. Sci. 44 (2006), no. 11-12, 738--747.

\bibitem[15]{Konieczny} Konieczny, P., On a steady flow in a three dimensional infinite pipe,
Coll. Math. 104 (2006), no. 1, 33--56.
	
\bibitem[16]{2dOseen} Konieczny, P., $L_p$-estimates for the Oseen system 
in 2D exterior domains, in preparation.
	
\bibitem[17]{Ladyz} Ladyzhenskaya, O.A.: The Mathematical Theory of Viscous
Incompressible Flow, Gordon and Breach, New York, 1966
			 		  
\bibitem[18]{Mucha1} Mucha, P.B., On the inviscid limit of the Navier-Stokes 
equations for flows with large flux, Nonlinearity 16 (2003), 1715--1732. 		
					  
\bibitem[19]{MuRa} Mucha, P. B.; Rautmann, R., Convergence of Rothe's scheme for 
the Navier-Stokes equations wish slip boundary conditions in 2D domains.
Z. Angew. Math. Mech. 86 (2006), no. 9, 691--701.

\bibitem[20]{Mucha5} Mucha, P. B., On a pump. Acta Appl. Math. 88 (2005), no. 2,
125--141.

\bibitem[21]{Solonnikov} Solonnikov, V.A.; Scadilov, V.E.: On a boundary  
value problem for a stationary system of Navier-Stokes equations, Trudy
Mat. Inst. Steklov. 125 (1973) 186--199



\end{thebibliography}
}
\end{document}